%% file: main.tex
\documentclass[submission,copyright,creativecommons]{eptcs}
\providecommand{\event}{QSQW 2020} 
\usepackage{amsmath,amsthm,amsfonts,amssymb,times,graphicx,color}
\usepackage{multirow}
\usepackage{graphicx}
\usepackage{subfigure}
\usepackage{mathtools}
\usepackage{url}
\usepackage{color}
\mathtoolsset{showonlyrefs}

\renewcommand{\top}{\intercal}

\include{PabMacros}

\title{Growing Random Graphs with Quantum Rules}
\author{Hamza Jnane
\institute{T{\'e}l{\'e}com Paris, LTCI, 19 Place Marguerite Perey, 91120 Palaiseau, France}
\email{hamza.jnane@telecom-paris.fr}
\and
Giuseppe Di Molfetta
\institute{Universit{\'e} publique, CNRS, LIS, Marseille, France and Quantum Computing Center, Keio University}
\email{\quad giuseppe.dimolfetta@lis-lab.fr}
\and
Filippo M. Miatto
\institute{T{\'e}l{\'e}com Paris, LTCI, 19 Place Marguerite Perey, 91120 Palaiseau, France}
\email{\quad filippo.miatto@telecom-paris.fr}
}

\def\titlerunning{Growing Random Graphs with Quantum Rules}
\def\authorrunning{H. Jnane, G. Di Molfetta \& F. M. Miatto}
\begin{document}
\maketitle

\begin{abstract}
Random graphs are a central element of the study of complex dynamical networks such as the internet, the brain, or socioeconomic phenomena.
New methods to generate random graphs can spawn new applications and give insights into more established techniques.
We propose two variations of a model to grow random graphs and trees, based on continuous-time quantum walks on the graphs.
After a random characteristic time, the position of the walker(s) is measured and new nodes are attached to the nodes where the walkers collapsed.
Such dynamical systems are reminiscent of the class of spontaneous collapse theories in quantum mechanics.
We investigate several rates of this spontaneous collapse for an individual quantum walker and for multiple non-interacting walkers.
We conjecture (and report some numerical evidence) that the models are scale-free.

\bigskip

\end{abstract}

\section{Introduction}
The intrinsic randomness of quantum physics can be exploited to generate ``truly random'' numbers.
If a system is prepared in a superposition of basis states, a measurement outcome in that basis is intrinsically random.
Such genuine randomness, which cannot be achieved with only classical means, has an important role in several fields such as cryptography \cite{shannon1949communication} and simulations \cite{metropolis1949monte}.

In this work we harness quantum randomness for generating random graphs.
Our motivation is that a large variety of real systems, from natural to socioeconomic phenomena \cite{strogatz2001exploring}, protein-protein interactions \cite{vazquez2003modeling}, the brain \cite{stam2007graph}, the internet \cite{wang2003complex}, financial trading \cite{tumminello2012identification}, and many other dynamical systems can be described by random graphs.
Moreover, the random collapse of the wave function of the quantum walker that is built into our models is reminiscent of the spontaneous collapse theories in quantum mechanics \cite{ghirardi1986unified}.

Although there are various ways to construct such discrete structures \cite{Barabasi, Erdos}, for most of them it is difficult to certify randomness based solely on the observed random graph states \cite{kolmogorov1998tables}.
On the other hand, the intrinsic randomness that emerges from measuring a quantum system in superposition of states can be used to give guarantees on the randomness of a graph.
Quantum walks on graphs are well-known \cite{aharonov2001quantum} and already lend themselves to many applications, including search algorithms \cite{childs2004spatial}, element distinctness \cite{ambainis2007quantum}  and isomorphism tests \cite{berry2011two}, discretization of differentiable surfaces \cite{aristote2020dynamical, arrighi2019curved} and cryptography \cite{vlachou2015quantum}.
Quantum walks are a natural choice in our models, and can be seen as the quantum counterpart of classic random walks which have been defined on all kinds of grids, in continuous and discrete time \cite{di2020quantum}.

Here we propose an original application of Quantum Walks on graphs to the problem of random graph generation.
In our models, the quantum system is given by one or more quantum walkers that evolve continuously on the graph.
Then, a measurement operation or a spontaneous collapse of the walker guides the evolution of the graph's adjacency matrix over time.  

We will present two models. The first can grow random trees and is based on a single quantum walk, while the second can grow general graphs and is based on two walkers.

The manuscript is organized as follows: we will first present our two models in sections II and III.
Then we will present a few numerical experiments where we grow random graphs and analyze various metrics and how they depend on the average collapse time.\\

To support the numerical explorations in this manuscript and to promote reproducibility, we have developed an open source python library \cite{miatto2020quantumgraphs}.

\section{Models}
\subsection{Single-walker model: random trees}
Let us consider an undirected graph $G = (V,E)$, where $V$ is the set of vertices, $V=\{0,...,N-1\}$, and $E$ is the set of the edges connecting the nodes.
Such graph can be described by a symmetric adjacency matrix $A_G$ of dimension $N\times N$, which has a 1 at position $(i,j)$ if the $i$-th node and the $j$-th node are directly connected by an edge, and 0 otherwise.
In our case, we use the adjacency matrix as the Hamiltonian of a continuously evolving system: the quantum walker.
The quantum walker is defined by its state on the nodes $\ket{\Psi_G} \in \mathbb{C}^N$ such that $\ket{\Psi_G} = \sum_{v\in V} a_v\ket{v}$, where $\ket{v}$ are the vertices of the graph.
The continuous time quantum walk is driven by the unitary operator $U(t) = e^{ - i A_G t}$. 
The probability of a walker starting at vertex $u$ to be measured at vertex $v$ after a time $t$ is given by
\begin{align}
P_t(v|u) = |\bra{u}U(t)\ket{v}|^2
\end{align}

The process for growing graphs can be summarized as follows:
\begin{enumerate}
\item set an initial graph state $|\Psi_G\rangle$ over the graph $G$;
\item let the walker evolve under the Hamiltonian $A_G$ for a time $t$ where $t \sim \mathrm{Exp}(\tau)$ is sampled from the exponential distribution of average $\tau$;
\item measure the position of the walker and obtain a node $v$ (this step is equivalent to sampling from the distribution $P(v) = |\langle v|U(t)|\Psi_G\rangle|^2$);
\item attach a new node to the node $v$, updating the adjacency matrix accordingly;
\item restart from 1.~with $|\Psi_G\rangle = |v\rangle$.
\end{enumerate}

Note that at each step the dimension of the Hilbert space associated to the graph $G$ increases by one.
As this model does not allow to create closed loops, it leads to the growth of random trees, so with vanishing cluster coefficient.

\begin{figure}[h!]
    \centering
        \includegraphics[width=0.2\textwidth]{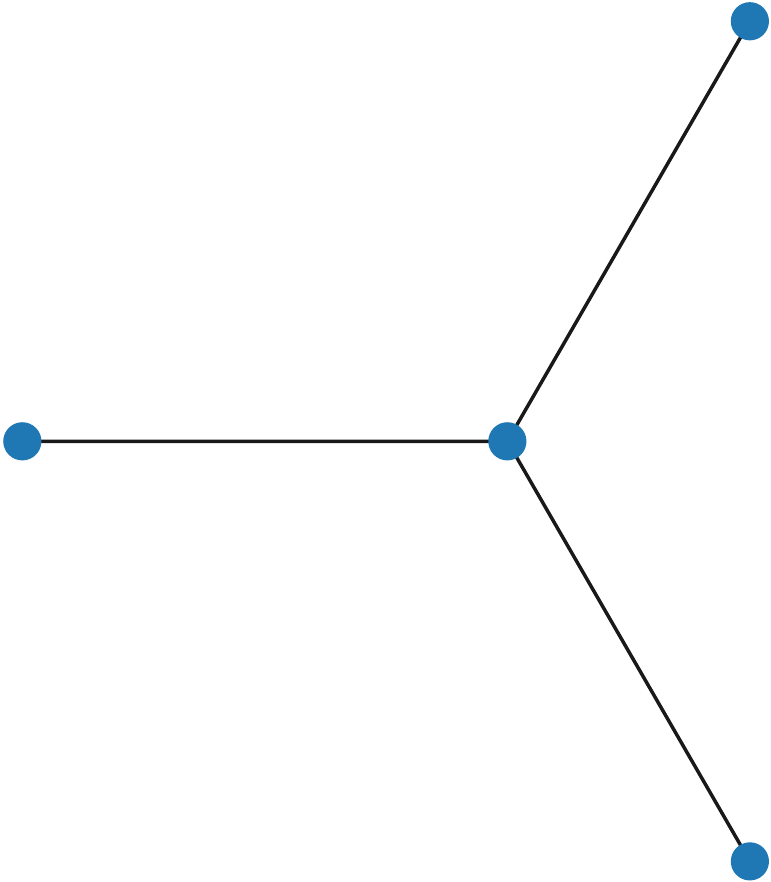}
        \includegraphics[width=0.3\textwidth]{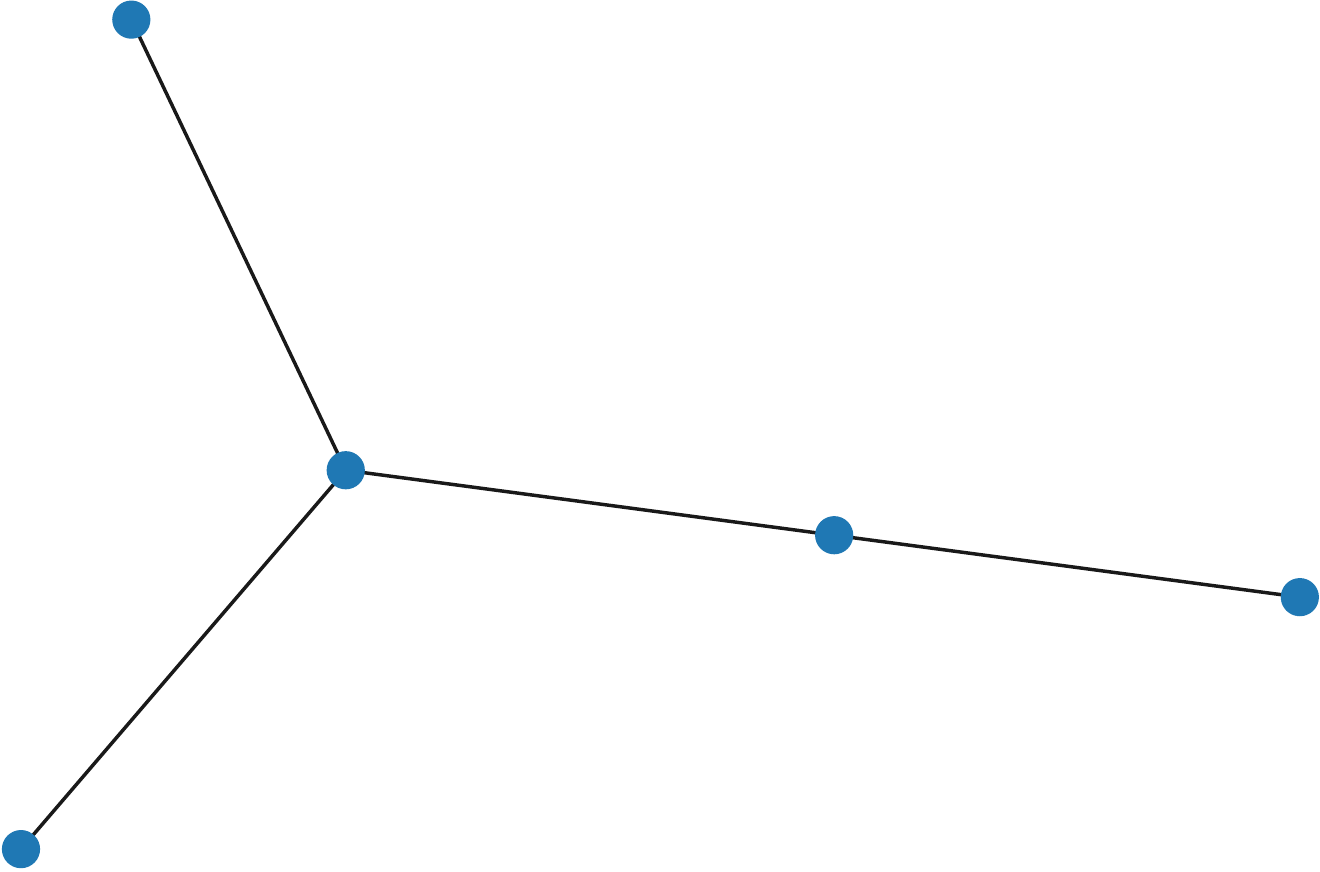}
        \caption{(left) Initial graph. (Right) Graph after 1 complete step}
        \label{fig:graph1}
\end{figure}

As an exemple, consider the graph represented by 
\begin{equation}
H = \begin{bmatrix} 0&1&1&1\\1&0&0&0 \\1&0&0&0 \\  1&0&0&0 \end{bmatrix}
\end{equation}
with the walker on the node $v=0$, i.e. with state $\ket{\Psi(0)} = (1, 0, 0, 0)^\top$.
Let the walker evolve for a time $t = 0.5$. The probability distribution of the walker's position is now $P(v) \approx  (0.42,  0.20,  0.19 , 0.19)^\top$. 
We measure the walker and we collapse its state, finding it for example on the node $v=1$.
Finally, as illustrated in Fig.\ref{fig:graph1}, we update the graph and the walker and we are set for a new iteration:
\begin{equation}
H' = \begin{bmatrix} 0&1&1&1&0\\1&0&0&0&1 \\1&0&0&0&0 \\  1&0&0&0&0 \\  0&1&0&0&0 \end{bmatrix} \qquad \qquad  |\Psi(\tau)\rangle =  \begin{bmatrix}
0 \\
1 \\
0 \\
0 \\
0
\end{bmatrix}
\end{equation}

\begin{figure}[ht!]
  \centering
  \includegraphics[width=0.2\textwidth]{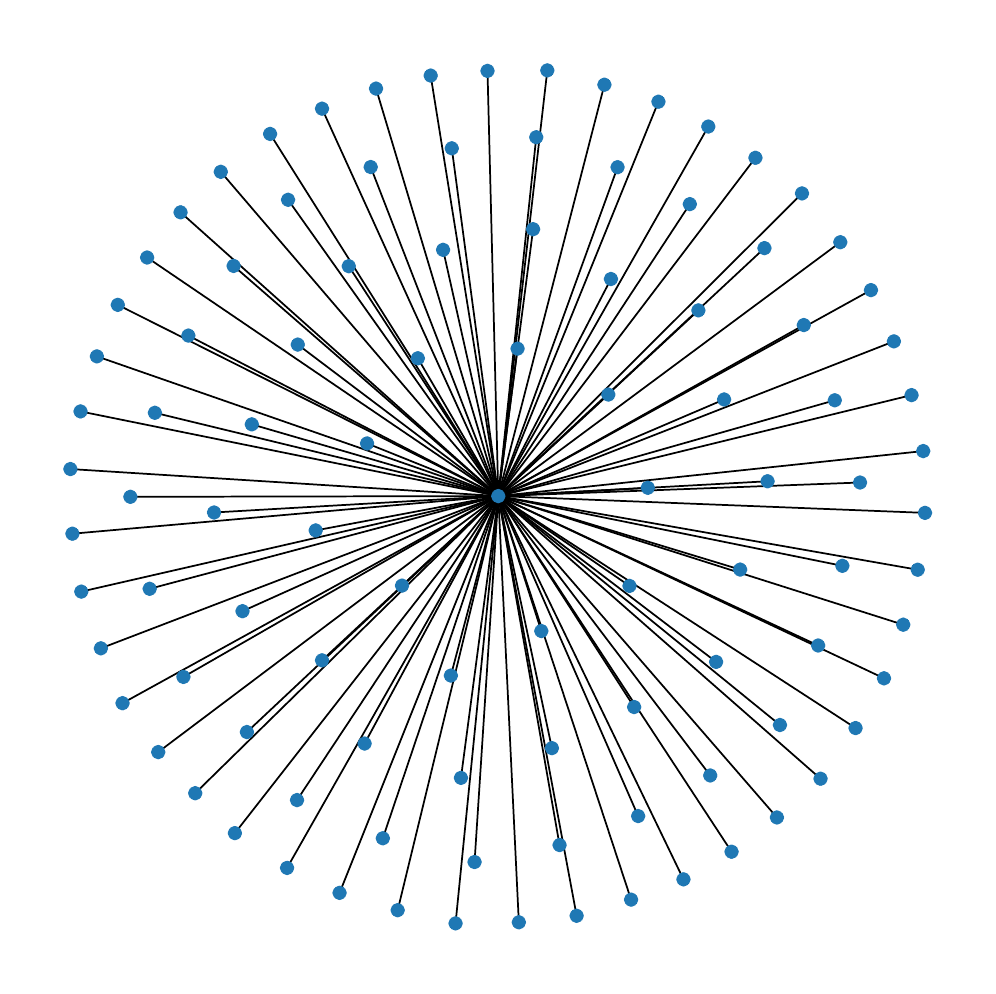}
     \includegraphics[width=0.20\textwidth]{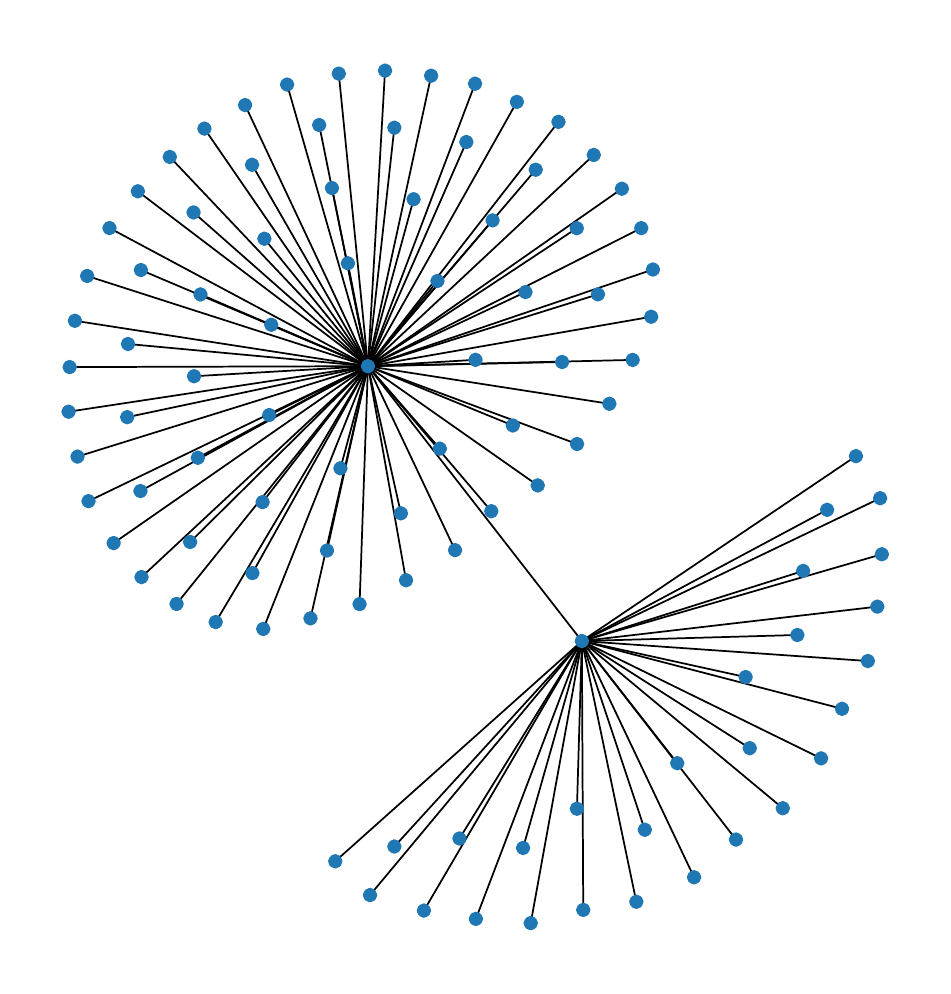}
   \includegraphics[width=0.20\textwidth]{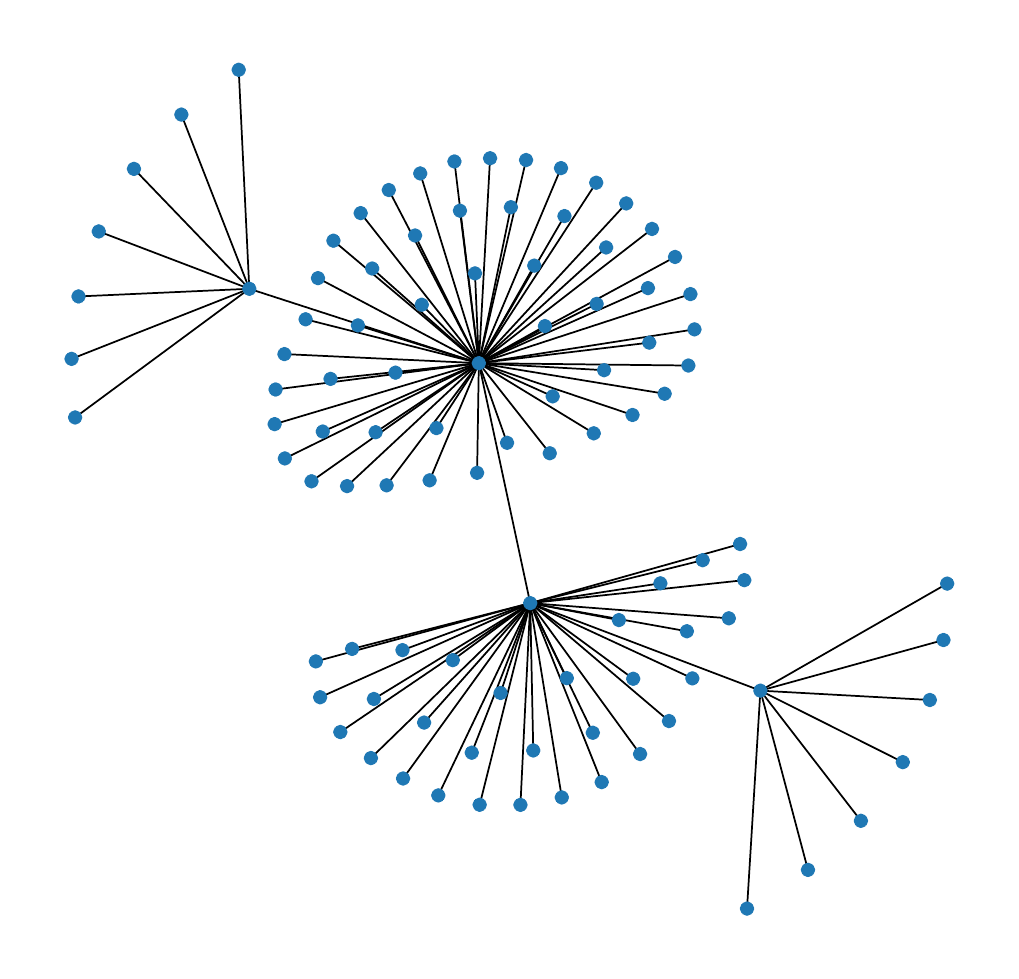}
      \includegraphics[width=0.22\textwidth]{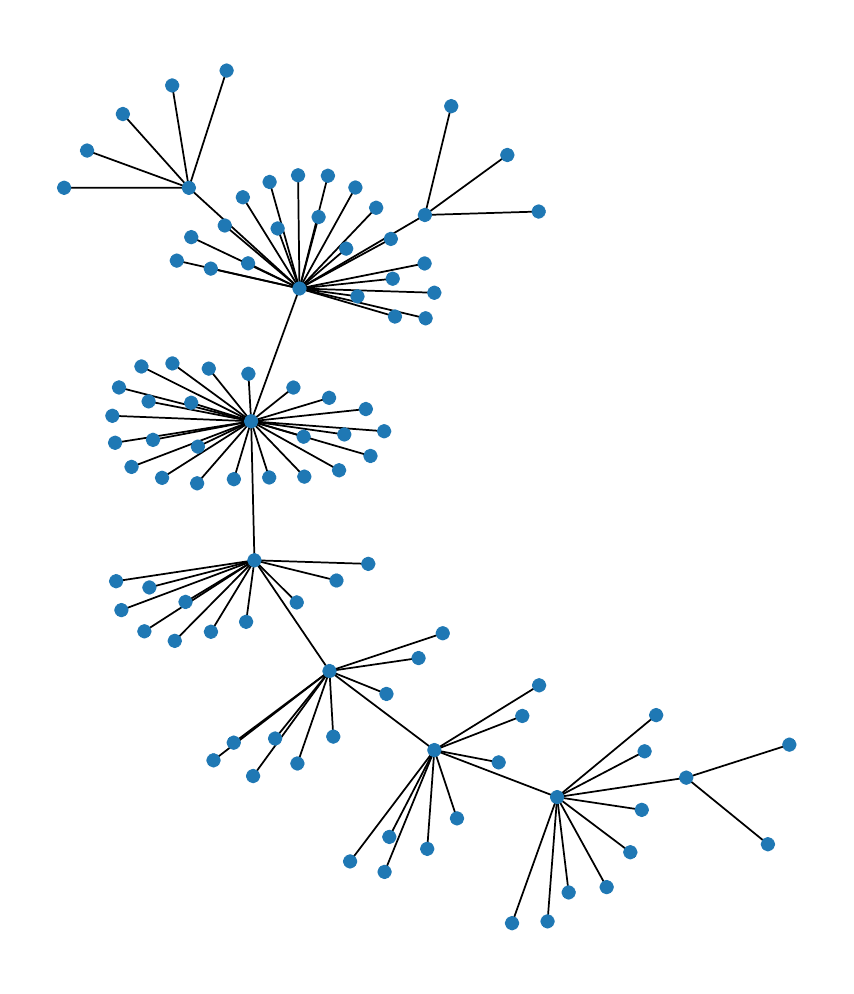}
       \includegraphics[width=0.20\textwidth]{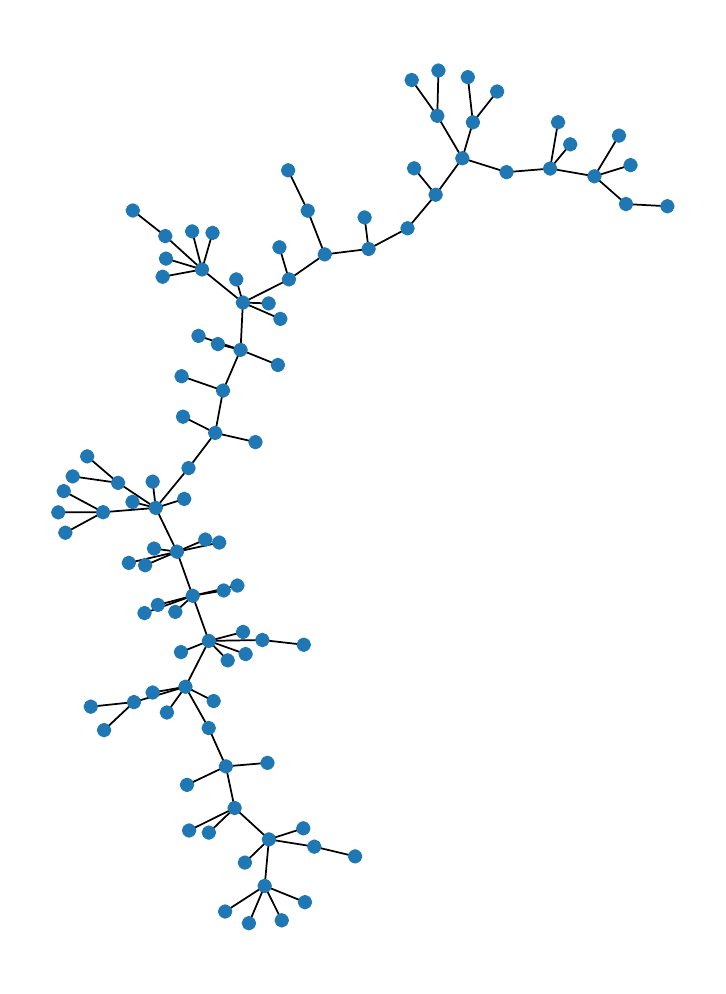}
        \includegraphics[width=0.26\textwidth]{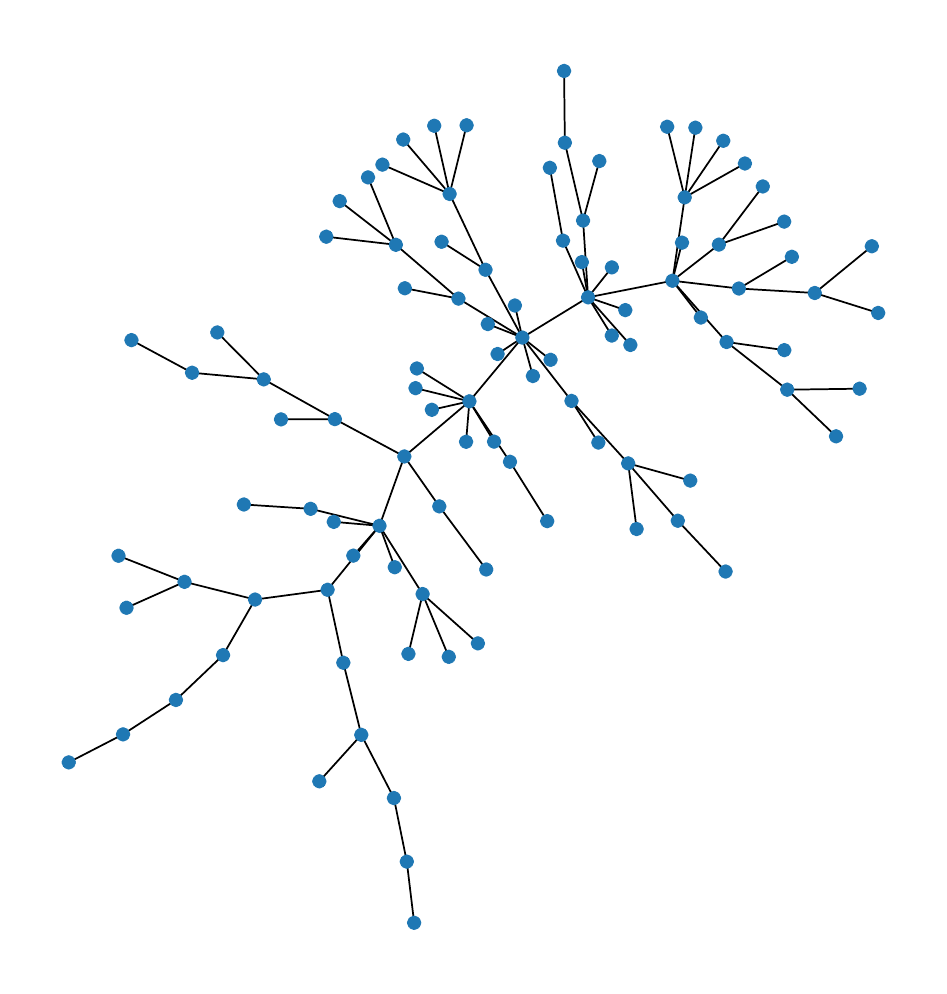}
          \caption{\label{fig:1walker}Generic random graphs for $n=100$ steps - one walker model for $\tau = 0.001, 0.01, 0.05, 0.1, 0.5, 10$.
          Notice how for longer average collapse times, the walker has more time to explore the graph, which leads to more complex trees.}
     \end{figure}

As illustrated in Fig. \ref{fig:1walker}, we observe that for short average time $\tau$ between collapses, the nodes aggregate in few large star graphs connected by single edges.
This can be understood by noticing that for short evolution times, the walker may not have enough time to leave the central node where it keeps collapsing.
However, as we show below, the dynamics in these sub-graphs is rich and indeed a walker is actually \emph{guaranteed} to leave a localised star graph because the rate at which the amplitude leaves the central node increases with the size of the star.

For longer average times instead the walker has time to explore the graph and the star sub-graphs don't have a way to develop.
It is an open question whether structures analogous to star graphs develop on a larger scale, leading to scale-free trees.

A walker that begins its evolution on the central node of a star graph keeps oscillating between the central node and an uniform superposition on the edges, which can be seen by analyzing the eigenvalues of the adjacency matrix of a star graph.
The frequency of oscillation for $n$ nodes is proportional to $\sqrt{n}$, and the probability to collapse on the central node after a time $t$ is given by $\cos^2(\sqrt{n}t)$.

Starting from a trivial graph with a single node, the probability to be in the same node at the center of the star for $k$ consecutive collapse events for a short average collapse time $\tau$ is
\begin{align} 
    P_\mathrm{center}(k) &\approx \prod_{n=1}^k \cos^2\left(\sqrt n\tau\right)\\
\end{align}
It's easy to see that $P_\mathrm{center}(k)$ decreases monotonically with $k$ for any value of $\tau$.
So the probability to escape the center after exactly $k$ events is
\begin{align} 
P_\mathrm{out}(k) &\approx  \sin^2(\sqrt{k}\tau)\prod_{n=1}^{k-1} \cos^2\left(\sqrt n\tau\right)
\end{align}

To first approximation, a star graph will cease growing as soon as the walker collapses to an outer node.
So we can compute the average star size by  
\begin{equation}
   \mathbb{E}[n]_\tau \approx \sum_{k = 1}^\infty k P_\mathrm{out}(k) \approx\frac{1}{\tau}
\end{equation}
And the expected number of stars for $N$ steps is approximately $N/\tau$.
Understandably, the longer we wait, the easier it is for the walker to escape smaller stars.
%

One way to study the structure properties of the graph is to study its spectrum.
To find how the spectrum evolves we compute the characteristic polynomial $\chi_n(x)$  of the adjacency matrix.
For example, a single star-graph is represented by the adjacency matrix
\begin{equation}
 A_n = \begin{bmatrix} 0&1&...&1\\1&0&...&0 \\ \vdots&&\ddots\\1&0&...&0 \end{bmatrix}.
 \end{equation}

The characteristic polynomial is then $ \chi_n(x) = \det(A_n-xI_{n}) = -x^{n-2}(x^2-n)$.
Hence, the eigenvalues are $0$ with multiplicity $n-2$ and $\pm \sqrt n$.
This in turn implies that there are two eigenstates that evolve with frequency $\sqrt{n}$ in opposite directions and several other ``frozen'' walker states on the outer edge of the star.

For $k$ stars, let $n_i$ be the number of nodes on the $i^{th}$ star, then:

$$ \chi_{n_1,...,n_k}(x) = -n_kx^{n_k-1}\chi_{n_1,...,n_{k-1}-1}(x) + x^{n_k}\chi_{n_1,...,n_{k-1}}(x) $$

\begin{figure}[h!]
    \centering
    \includegraphics[width= 0.3\textwidth]{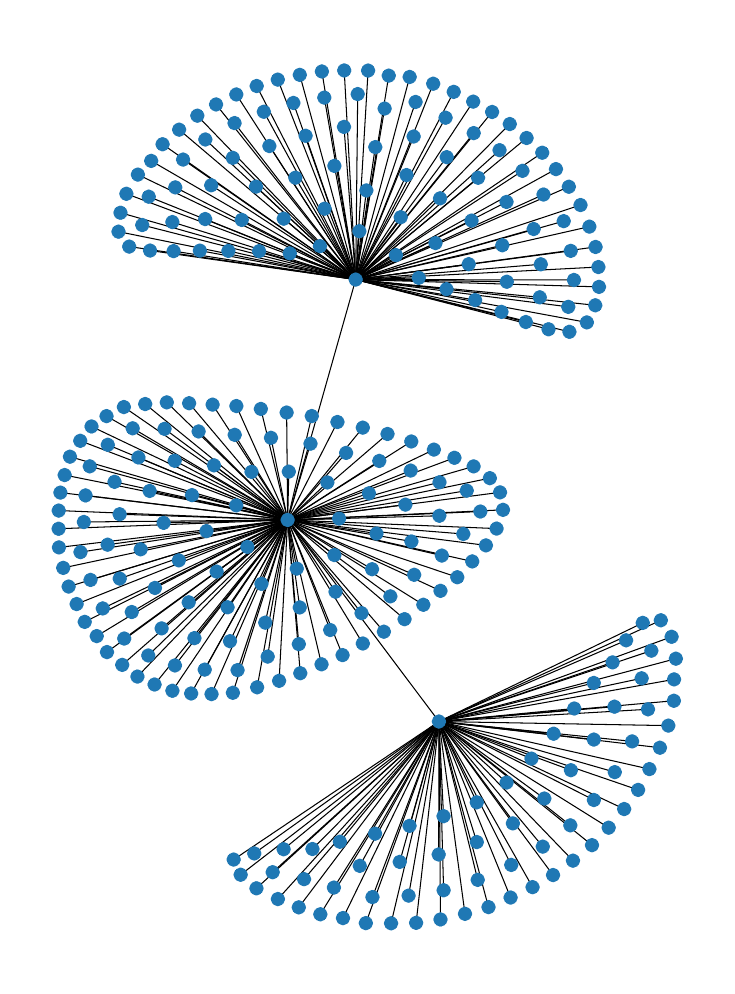}
    \includegraphics[width= 0.46\textwidth]{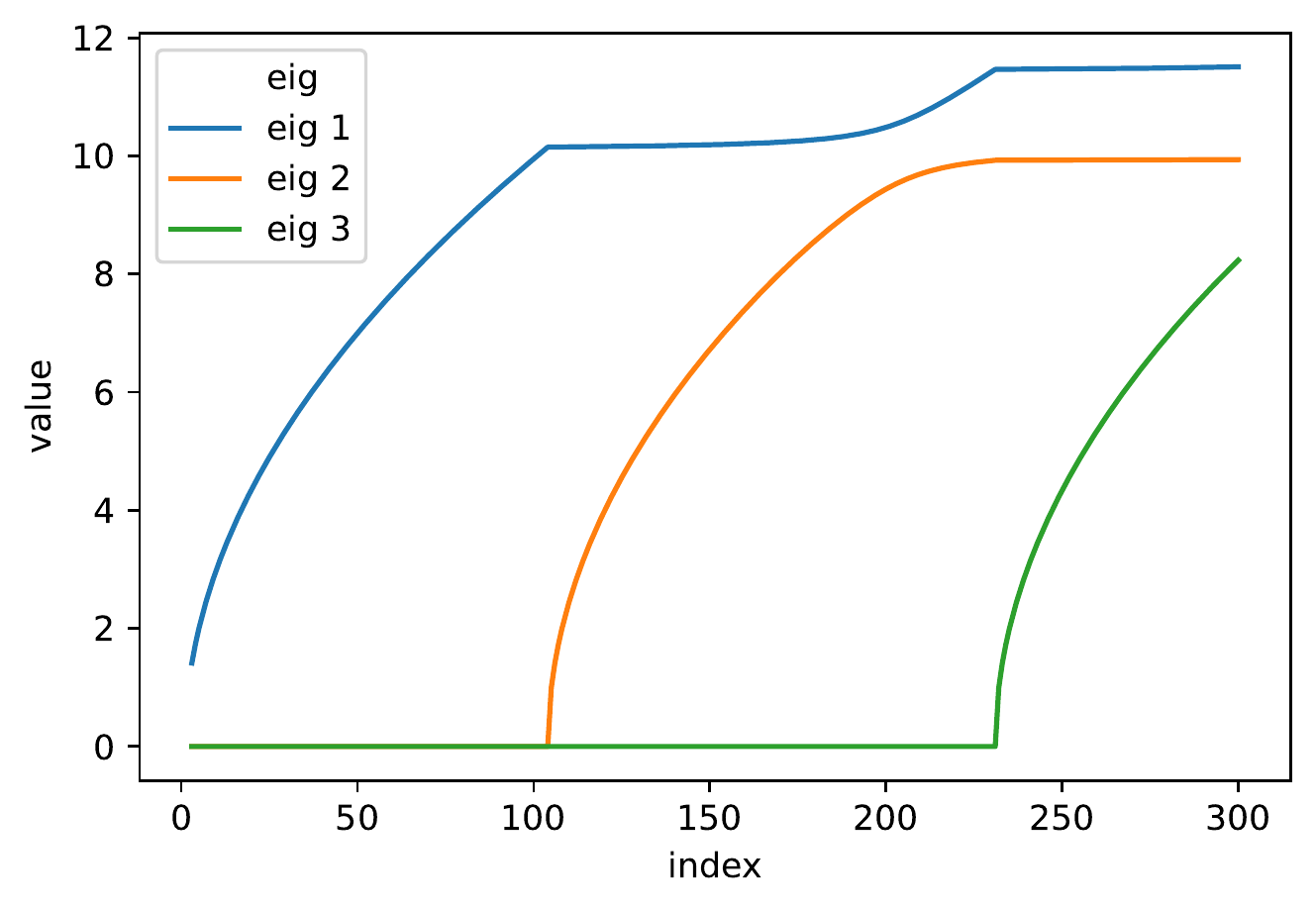}
    \caption{\label{fig:eval1}
    Top: A random graph that evolved into three connected star graphs.
    We expect therefore three nonzero eigenvalues (modulo sign) representing the oscillation frequency of walkers that begin at the center of the stars.
    Bottom: Evolution of the eigenvalues.
    As soon as a new star graph begins being populated by new nodes, the corresponding new frequencies also increase.
    At the same time, the growth of the previous eigenvalues stops.
    (Blue: first eigenvalue, orange: second eigenvalue, green: third eigenvalue.)}
\end{figure}

We finally get the spectrum by finding the roots of these polynomials and we get the same evolution as we can see in Fig.\ref{fig:eval1}.

\subsection{Multiple walkers model: random graphs}

\begin{figure}[ht!]
    \centering
    \includegraphics[width=0.2\textwidth]{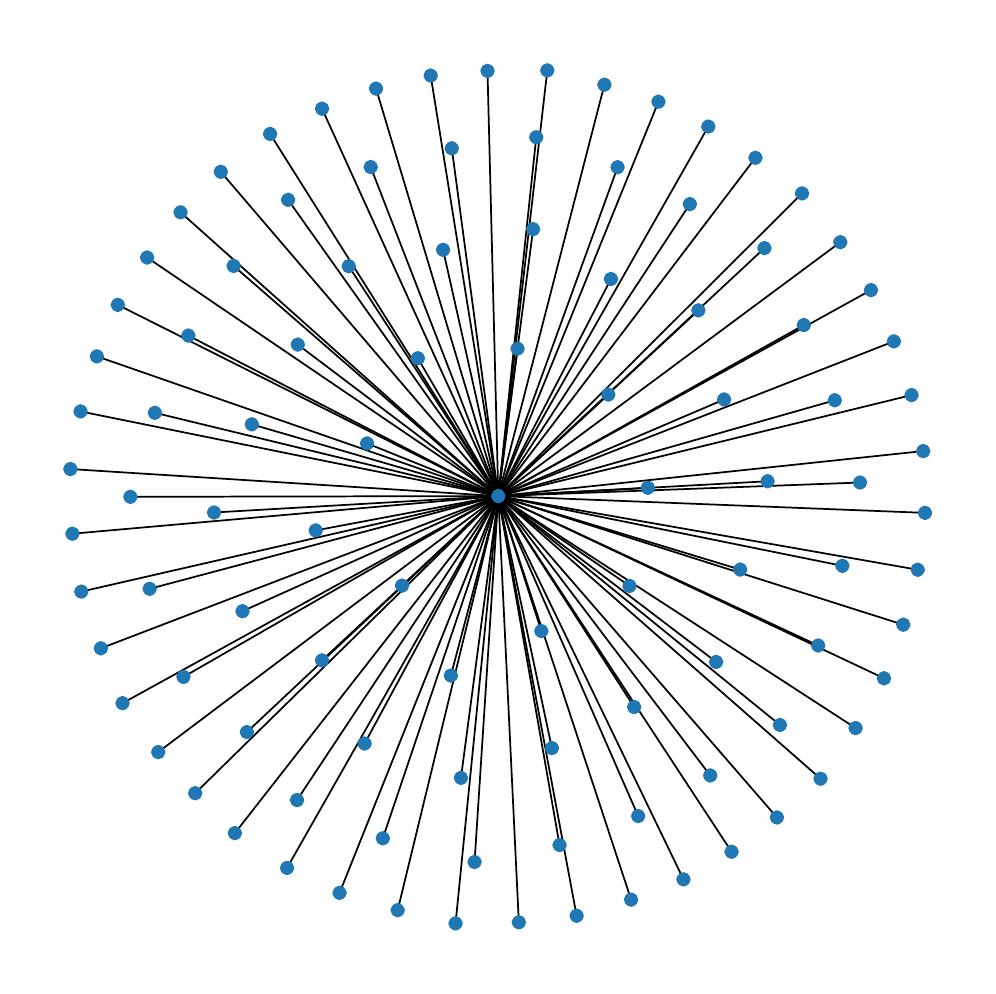}
    \includegraphics[width=0.20\textwidth]{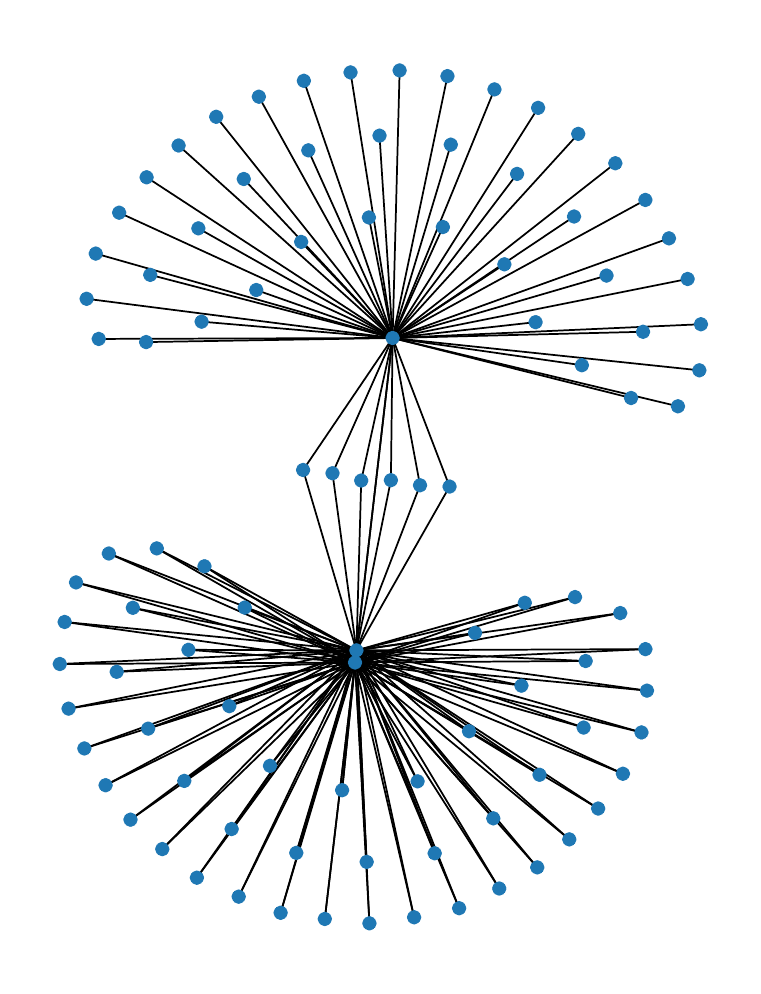}
  \includegraphics[width=0.20\textwidth]{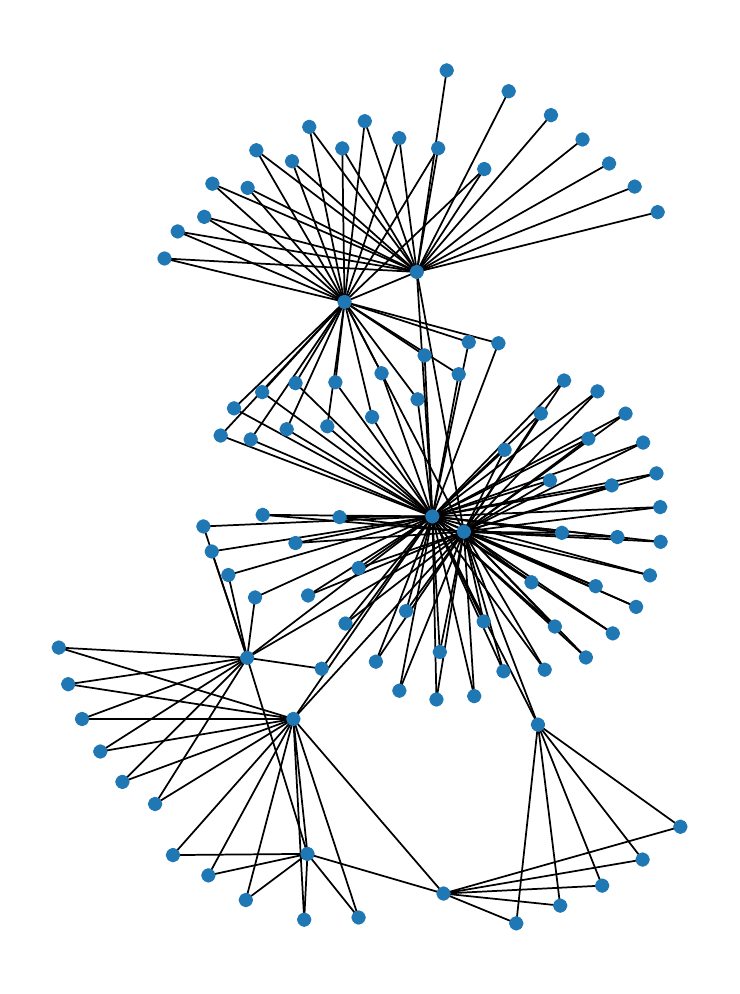}
     \includegraphics[width=0.22\textwidth]{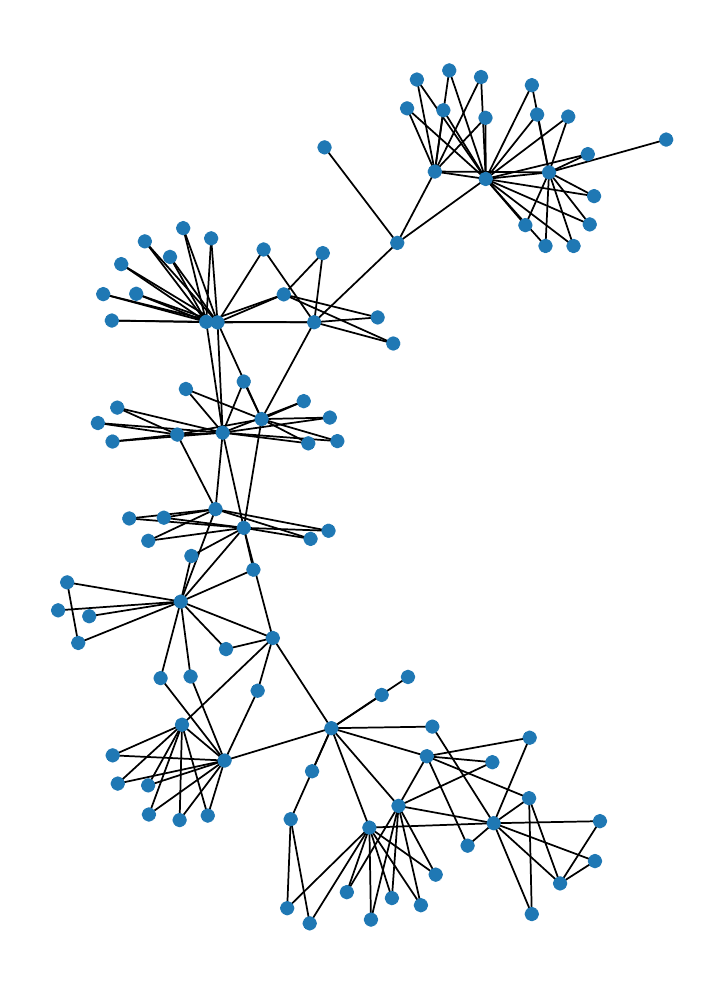}
      \includegraphics[width=0.20\textwidth]{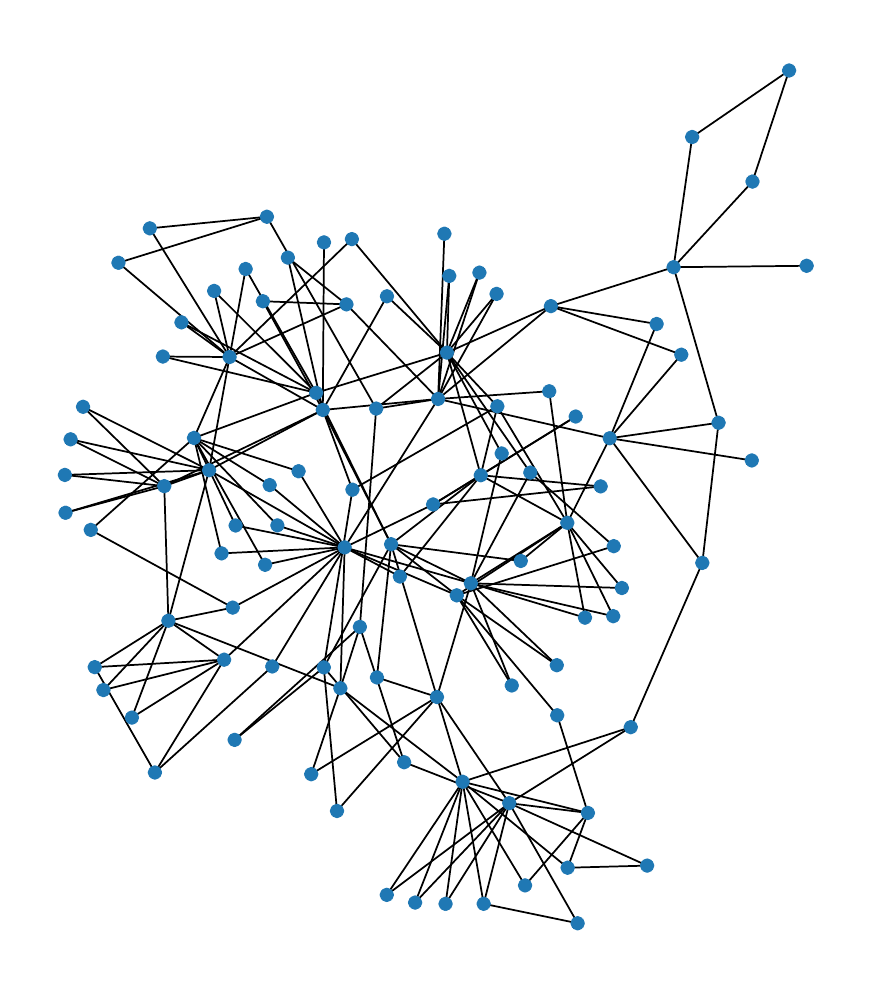}
       \includegraphics[width=0.26\textwidth]{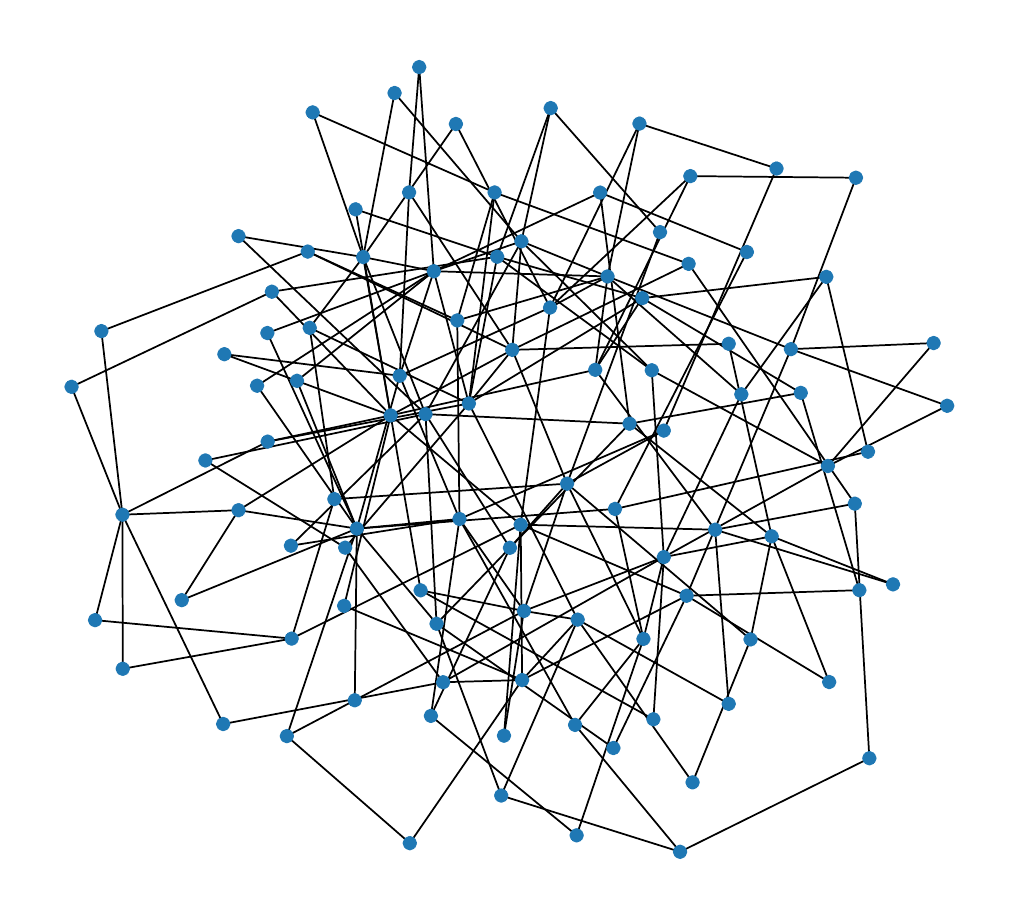}
            \caption{\label{fig:2walker}Generic random graphs for $n=100$ steps - two walker model for $\tau = 0.001, 0.01, 0.05, 0.1, 0.5, 10$.
            As in the single-walker case, longer exploration times lead to more connected random graphs.}
       \end{figure}

In order to build structures with a non-zero cluster coefficient, we can use two or more walkers.
In this model we have two independent walkers that collapse at the same time. We attach a new node to both the nodes where the walkers collapsed.
This allows closed loops to form, thus with non-vanishing cluster as we can see in Fig.\ref{fig:2walker}.
If the walkers happen to collapse to the same node, the new node will have a single edge.


\section{Qualitative analysis}
%

\begin{figure}[h!]
    \centering
        \includegraphics[width=0.5\textwidth]{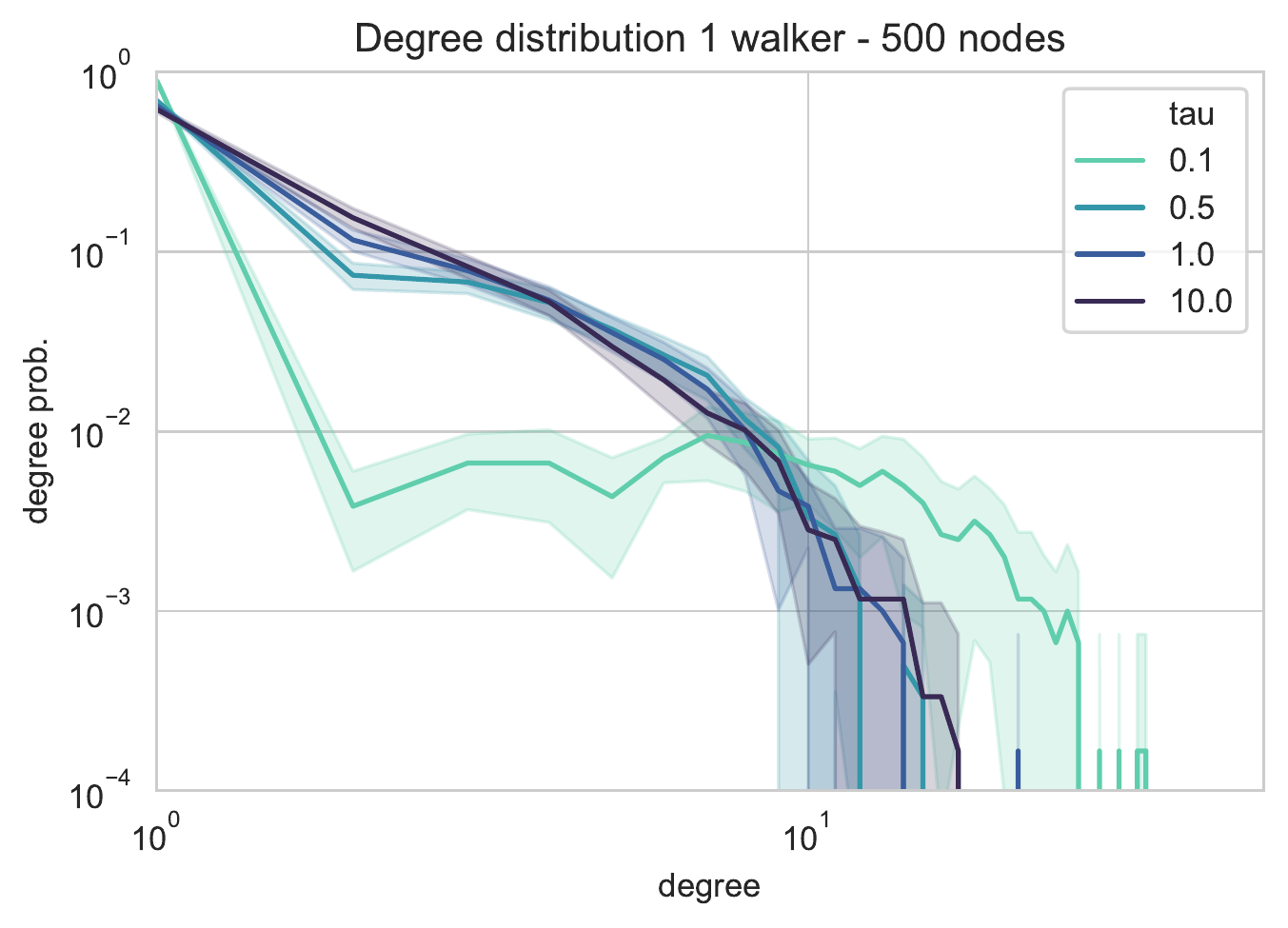} 
         \includegraphics[width=0.5\textwidth]{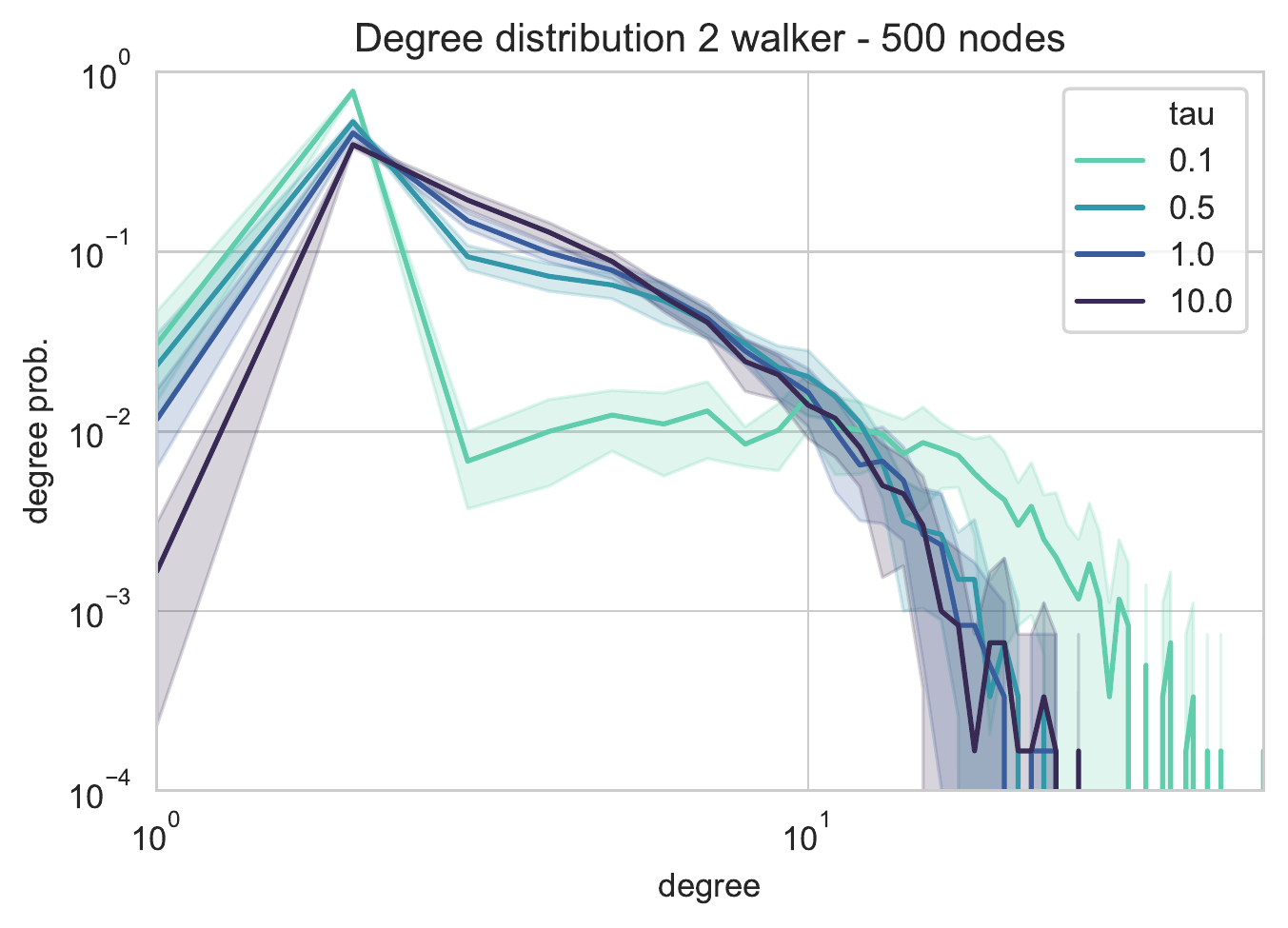}
         \includegraphics[width=0.5\textwidth]{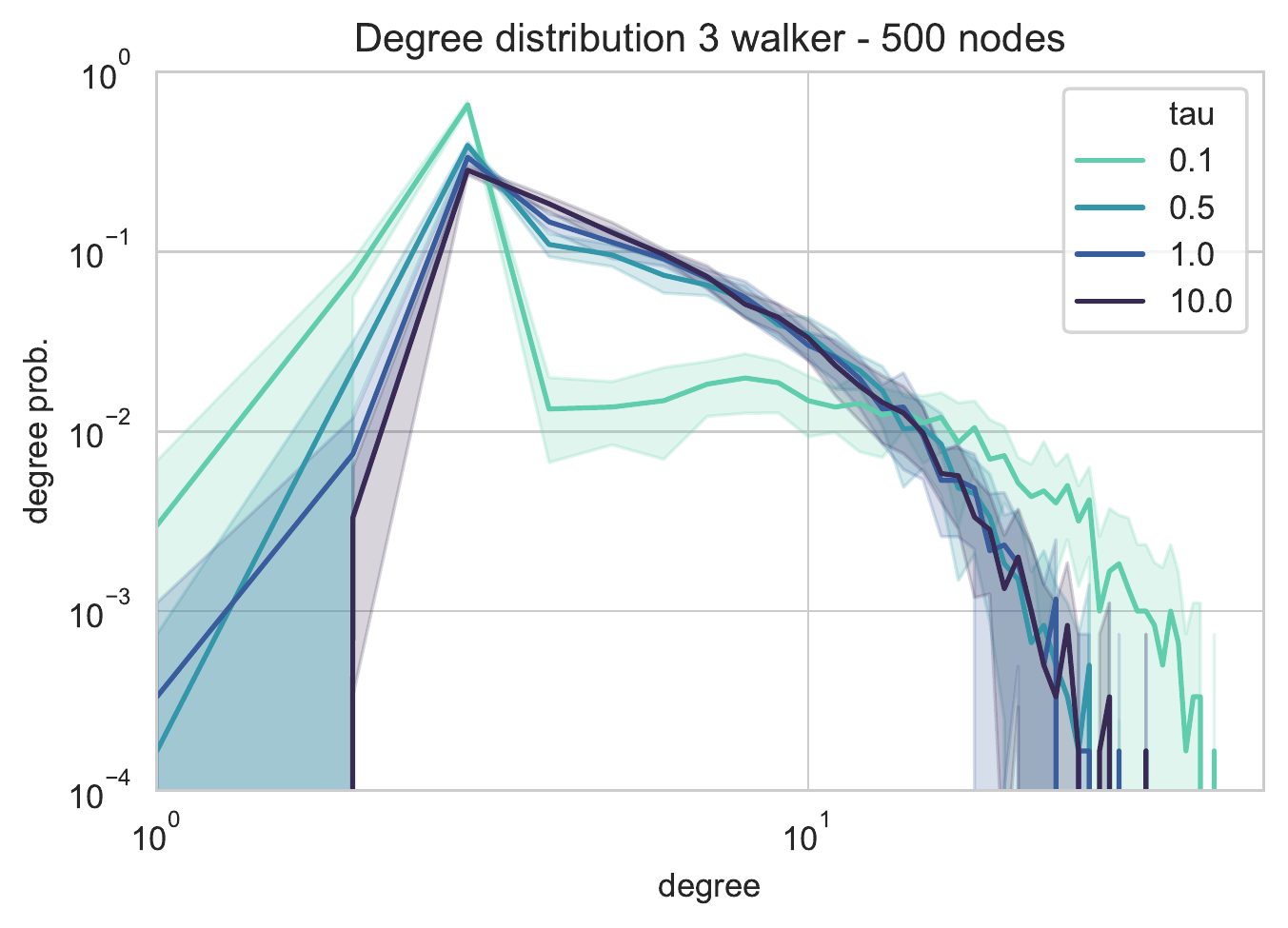}
        \caption{\label{fig:degree}Degree distribution for random graphs of $n = 500$ nodes.
        As the value of the exploration parameter increases, the random graphs seem to reach a scale-free structure for large $\tau$.}
\end{figure}

We first look at the degree distribution. For each degree $k$ we compute the fraction $d(k)$ of nodes on the graph that has $k$ connections.
By studying the degree distribution, we can get an indication of whether our models generate scale-free networks i.e. whose degree distribution follows a power-law $d(k) = k^{-\alpha}$.
We compute the average degree distribution for the first model, with 1000 and 100 nodes.
We find that the degree distribution seems to converge, for large $\tau$, to a power law, as shown in Fig.~\ref{fig:degree}.

\begin{figure}[h!]
    \centering
     \includegraphics[width=0.5\textwidth]{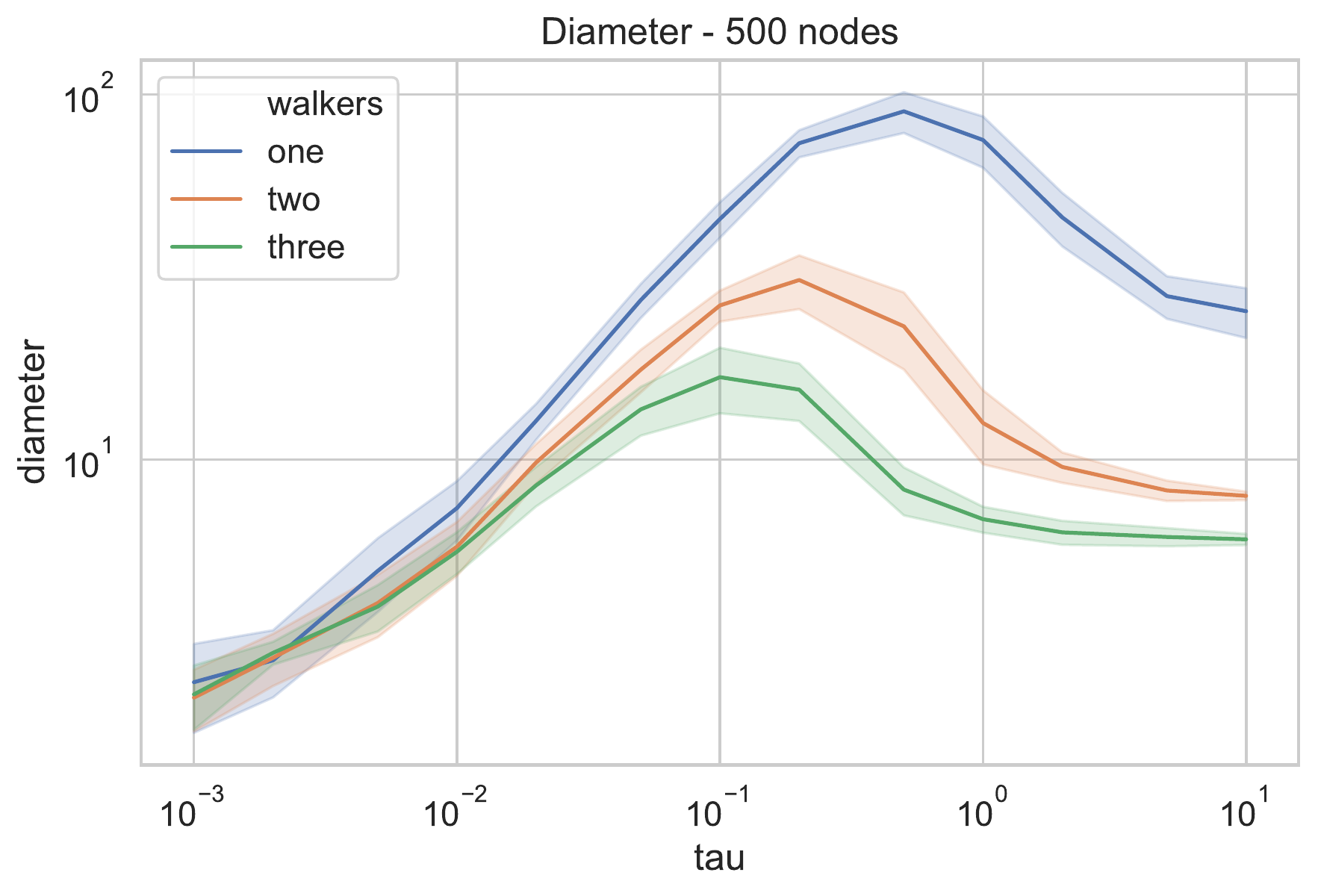}
        \caption{\label{fig:diameter}Diameter of the graph for the single-walker model for 100 and 1000 nodes.}
    \end{figure}

Next we turn to the diameter of the graphs.
To compute the diameter of a graph we compute the maximum value of the shortest path between every pair of edges.
As we can see in Fig.~\ref{fig:diameter}, the diameter dependence on time depend on the number of walkers and is maximal for $0.1 < \tau < 1$. 

\begin{figure}[h!]
    \centering
     \includegraphics[width=0.5\textwidth]{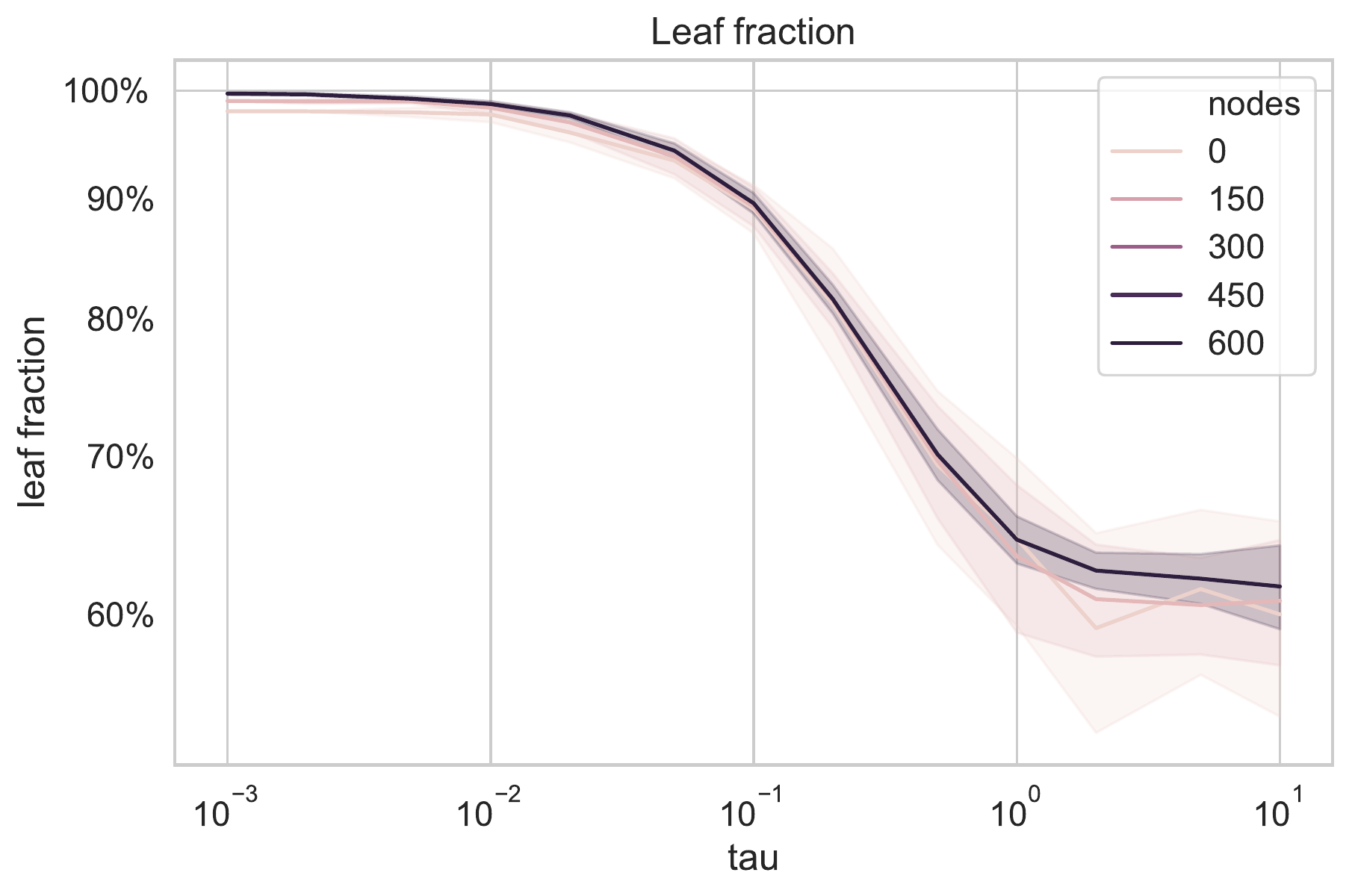}
        \caption{\label{fig:leafs}Number of nodes on the edge of the star for the Single-walker model.
        As the leaf fraction has the same dependence on $\tau$ for various graph sizes, it is very likely that it constitutes an invariant property.
        There also seems to be a phase transition between two different regimes.}
\end{figure}

We observe an interesting phenomenon when we compute the fraction of nodes that live on the edge of the graphs.
We define this as the ``leaf fraction'', i.e. the fraction of nodes with only one neighbor.
As can be seen in Fig.~\ref{fig:leafs}, the leaf fraction decreases with $\tau$ until it becomes stable from around $\tau\approx 1$.
This effect seems to be independent of the size of the graphs, and we conjecture that is a universal property of this model.

Finally, we compute the average clustering coefficient.
The local clustering coefficient of a node is a measure of how much its neighbors are linked to each other.
It is a coefficient introduced by Watts and Strogatz in \cite{Watts} to find whether a graph exhibits the small-world property.

To define the clustering coefficient, we need to define what is the neighborhood of a node in a graph. 
The neighborhood $N_i$ of a node $i$ is defined as the set of all its neighbors. Let $G=(V,E)$ be a graph of $n$ nodes with $V = [\![ 1,n ]\!]$ and $E$ the set of couples $(i,j)$ if there's a link between the nodes $i$ and $j$.
Then we define the local clustering coefficient of the node $i$ which has $n_i$ neighbors as $C_i = \frac{2|\left\{(j,k) \; j,k \in N_i, (j,k) \in E\right\}|}{n_i(n_i-1)}$ for an undirected graph.
Finally, we obtain the graph's average clustering coefficient : 
$C = \frac{1}{n}\sum_{i=1}^n C_i$. We can notice that the clustering coefficient decreases rapidly for large $\tau$ and seems to be independent on the number of walkers, with the only exception of the single-walker model. 

\begin{figure}[h!]
        \centering 
        \includegraphics[width=0.5\textwidth]{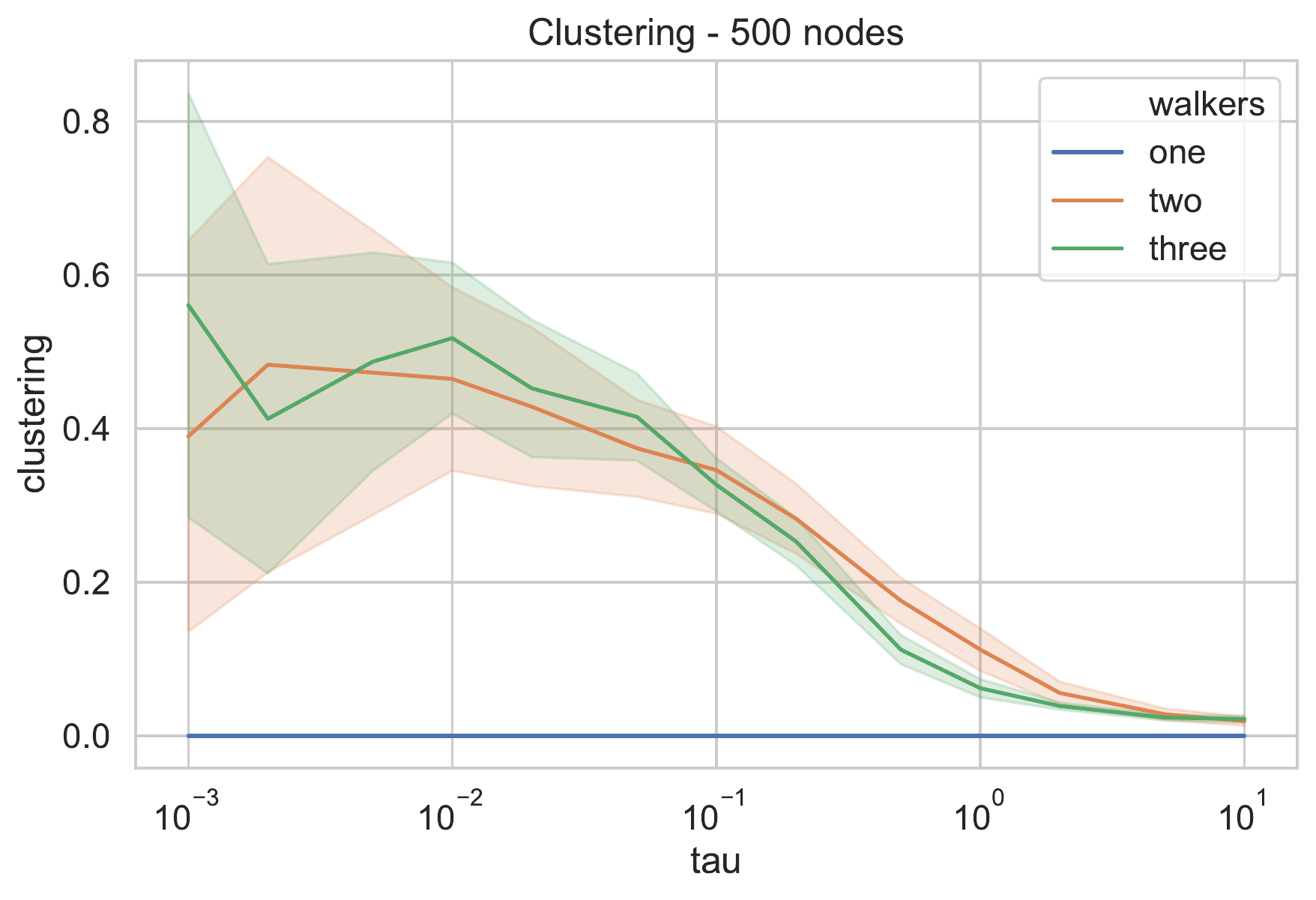}
        \caption{Clustering coefficient of one-, two- and three-walkers respectively.
        The one-walker model can only produce random trees and therefore the clustering coefficient is always zero.}
  \end{figure}

\section{Discussion}
We have presented two new models for growing random graphs, that are driven by the continuous evolution of quantum walkers and by spontaneous collapses of their wave function.
We have numerical evidence that the random graphs that result from these models display a power-law degree of distribution, which indicates a scale-free structure.

We also have found two interesting phenomena in the average graph diameter and in the fraction of nodes that live on the edge.
An average collapse time of $\tau\approx 1$ seems to play a special role, while for $\tau \gg 1$ it seems to lead the measures to relax to an asymptotic value.
In particular, the leaf fraction is a quantity that seems to give us information about the bulk of a graph from its surface, which is reminiscent of the holographic principle.
We suspect that around $\tau\approx 1$ there could occur a first order phase transition, as Fig.~\ref{fig:diameter} seems to suggest.

We note that as the wave function of the quantum walker can interfere with itself, the continuous-time quantum walk does not tend to a limiting distribution (which instead is characteristic of classical random walks).
This might result in fundamental differences between classical and quantum models that are not possible to reconcile.
Moreover, we note that as the off-diagonal elements of the adjacency matrix are never negative, the Hamiltonian that it represents is ``stoquastic'' \cite{bravyi2010complexity} and therefore the evolution of the walker should not be exponentially hard to compute.
Because of this, it could be that even if these growth models were not reproducible by classical random walk models, they could be reside in an efficient class of computational complexity.

Among the directions of future work one could introduce an additional degree of freedom in terms of a quantum coin, similarly to what is done in discrete-time quantum walks.
Another possible perspective is to explore the possibility to perform quantum search algorithms over growing random graphs, for example by combining multiple walkers where e.g. two of them are in charge of growing the graph and a third performs the search while preserving its unitarity.
Our hope is that this will produce ``optimized'' random graphs for the objective that the walker is searching for. 

\nocite{*}
\bibliographystyle{eptcs}
\bibliography{Biblio.bib}
\end{document}

%% file: PabMacros.tex
\newcommand{\couic}[1]{}
\newcommand{\couicfootnote}[1]{}
\newcommand{\couicefootnote}[1]{}

\newcommand{\ket}[1]{| #1 \rangle}
\newcommand{\bra}[1]{\langle #1 |}